\documentclass{elsart}

\usepackage{epsfig}

\usepackage{amssymb,amsmath}

\usepackage{url}

\begin{document}

\begin{frontmatter}



\title{Photon helicity in $\boldsymbol{\Lambda_b \rightarrow pK\gamma}$
decays}


\author{Federica~Legger,}
\author{Thomas~Schietinger\thanksref{PSI}}

\address{Laboratory for High-Energy Physics,
Ecole Polytechnique F\'ed\'erale,\\ CH-1015 Lausanne, Switzerland}

\thanks[PSI]{Now at Paul Scherrer Institut, CH-5232 Villigen PSI, Switzerland}

\begin{abstract}
Radiative decays of polarized $\Lambda_b$ baryons represent an
attractive possibility to measure the helicity of the photon emitted
in the $b\rightarrow s\gamma$ quark transition and thus to subject
the Standard Model to a stringent test at existing and future hadron
colliders. The most abundant mode, $\Lambda(1116)\gamma$, is
experimentally very challenging because of the long decay length of
the $\Lambda(1116)$. We show that the experimentally more accessible
$\Lambda_b \rightarrow pK\gamma$ decays proceeding via $\Lambda$
resonances may be used to extract the photon helicity for sufficient
$\Lambda_b$ polarization, if the resonance spin does not exceed 3/2.
A direct comparison of the potential of such resonance 
decays to assess the photon polarization at a hadron collider with 
respect to the decay to $\Lambda(1116)$ is given.
\end{abstract}

\end{frontmatter}


\section{Introduction}

The helicity of the photon emitted in the $b\rightarrow s\gamma$
quark transition remains one of the last untested predictions of the
Standard Model (SM) in the realm of $B$ physics
\cite{Atwood-Gronau-Soni}. Given the experimental difficulty of
directly measuring the helicity of the emitted photon, several
indirect methods for its determination in $B$ meson decays have been
proposed, implying $B$-$\overline{B}$ interference
\cite{Atwood-Gronau-Soni}, photon conversion to $e^+e^-$
\cite{conversion}, resonant states in the $K\pi\pi^0$ final state
\cite{GGPR}, and interference with radiative charmonium decays
\cite{Knecht-Schietinger}. A particularly attractive possibility
arises from the decay of $b$-baryons, as first pointed out by Gremm,
Kr\"uger and Sehgal \cite{GKS}, and further elaborated on by Mannel
and Recksiegel \cite{Mannel-Recksiegel}. A rather complete study of
polarized $\Lambda_b \rightarrow \Lambda\gamma$ decays in the
context of a high-luminosity $Z$ factory (``Giga-$Z$'') has been
given by Hiller and Kagan \cite{Hiller-Kagan}.

Despite considerable efforts, the results obtained at the $e^+e^-$
$B$ factories do not yet put significant constraints on the photon
polarization in $b\rightarrow s\gamma$ \cite{Moriond}. With the
Giga-$Z$ factory relegated to a distant future, but dedicated $B$
experiments at hadron colliders imminent, it is worth considering
the potential of polarized $\Lambda_b$ decays at hadron colliders.
At the Large Hadron Collider (LHC), for instance, the $\Lambda_b$
baryons produced in $pp$ collisions are expected to be polarized
transversally with respect to the production plane, with a
polarization possibly as large as 0.2 \cite{Ajaltouni}. This
polarization can be measured with an estimated statistical precision
of 0.01 by an angular analysis of the decay $\Lambda_b\rightarrow
\Lambda J/\psi$ \cite{Hrivnac}.

An experimental issue arising at hadron colliders is the macroscopic decay
length of the $\Lambda$ baryon ($c\tau$ = 7.89 cm \cite{PDG04}).
It typically escapes the innermost parts of a large detector system without
leaving any trace, before weakly decaying, predominantly into a nucleon and a pion.
This poses a severe problem to experiments relying on the observation of a decay
vertex detached from the primary vertex to identify events containing $b$-hadrons
\cite{CDF-trigger,LHCb-trigger}, since neither the photon nor the
$\Lambda$ baryon from the $\Lambda_b \rightarrow \Lambda\gamma$ decay produce a
suitable signature.
A possible way around this problem is afforded by considering radiative $\Lambda_b$
decays to $\Lambda$ resonances above the nucleon-kaon ($N\overline{K}$) threshold,
such as $\Lambda(1520)$ or $\Lambda(1670)$.
With their prompt decay into $pK^-$ these resonances trace back the decay of
the $\Lambda_b$, thus rendering it more accessible to the online and offline
event selection.

The purpose of this Letter is to investigate the potential of
$\Lambda_b\rightarrow \Lambda(X) \gamma$ decays ($X$ = 1520, 1670, 1690, \ldots) for
assessing the photon polarization in the $b\rightarrow s\gamma$ transition at
hadron colliders, in particular in comparison to $\Lambda_b\rightarrow \Lambda(1116)
\gamma$.

\section{Photon polarization parameters}

Decays of the type $\Lambda_b\rightarrow \Lambda \gamma$ are
mediated by the quark transition $b\rightarrow s \gamma$. Long
distance contributions, arising from $W$ or intermediate meson
exchange, have been found to be negligible \cite{Mannel-Recksiegel}.
In the usual framework of an effective Hamiltonian, the relevant
operators contributing at leading order (LO) in $\alpha_s$ are the
electromagnetic dipole operators $O_7\mbox{}^{(}\mbox{}'\mbox{}^{)}
= (e m_b)/(16\pi^2)\overline{s} \sigma_{\mu\nu} R(L)b F^{\mu\nu}$,
responsible for the emission of a left- or right-handed photon, respectively:
\begin{equation}\label{eq:Heff}
    \mathcal{H}_\mathrm{eff} = -4\frac{G_F}{\sqrt{2}}V^*_{ts} V_{tb} (C_7 O_7+C'_7 O'_7),
\end{equation}
with $G_F$ the Fermi constant and $C_7\mbox{}^{(}\mbox{}'\mbox{}^{)}$ the Wilson
coefficient of the local operator $O_7\mbox{}^{(}\mbox{}'\mbox{}^{)}$;
$V_{tb}$ and $V_{ts}$ are the Cabibbo-Kobayashi-Maskawa (CKM) matrix elements.
In the operator definition $e$ is the electric charge, $m_b$ the mass of the $b$-quark,
$F^{\mu\nu}$ the electromagnetic field tensor and $\sigma_{\mu\nu} =
\frac{i}{2}(\gamma_\mu\gamma_\nu - \gamma_\nu\gamma_\mu)$.
$R = (1+\gamma_5)/2$ and $L = (1-\gamma_5)/2$ are the right- and left-handed
projectors, respectively.
Thus in leading order, the occurrence of right-handed photons is given
by the ratio of Wilson coefficients, $r = C_7'/C_7$.
In the SM, $r = m_s/m_b$ by virtue of the chirality of the $W$ exchanged
in the decay loop \cite{Atwood-Gronau-Soni}.
Various scenarios beyond the SM such as left-right symmetric models
predict new contributions to $C_7'$ and therefore larger values for $r$.
Hence the strong interest in constraining $r$ experimentally.

From the experimental point of view, the observable of interest is the
photon asymmetry
\begin{equation} \label{eq:alpha}
  \alpha_{\gamma}= \frac{P(\gamma_L) - P(\gamma_R)}{P(\gamma_L) + P(\gamma_R)},
\end{equation}
where $P(\gamma_{L(R)})$ represents the probability of producing
a left-(right-)handed photon in the decay.
In the leading-order limit, where only $O_7$ and $O_7'$ contribute,
$\alpha_{\gamma}$ is related to $r$ by
\begin{equation} \label{eq:alphaLO}
    \alpha_{\gamma}^\mathrm{LO} = \frac{1-|r|^2}{1+|r|^2} .
\end{equation}
Recently, however, it was shown that gluon bremsstrahlung contributions to
the matrix elements of operators other than $O_7$ and $O_7'$ can give significant
contributions to $\alpha_{\gamma}$, such that an experimental determination
of that asymmetry would only yield an effective ratio, $r_\text{eff}$ \cite{GGLP}.
In the following, we will set $r_\text{eff} \equiv r$ to simplify our
notation, but keep in mind that the relation between $\alpha_\gamma$ and
$r$ may be more complicated.
Furthermore, we assume that $r$ is the same for $\Lambda_b$ and
$\overline{\Lambda}_b$ decays, i.e., we do not consider CP violating effects.

\section{Properties of $\boldsymbol{\Lambda}$ resonances}

In Table~\ref{tab:lambdares} we list the properties of the better
known $\Lambda(X)$ resonances, many of which can only be crudely
estimated at this time, based on data compiled by the Particle Data
Group \cite{PDG04}. The Table also lists our estimates for the
branching fractions for the decays
$\Lambda_b\rightarrow\Lambda(X)\gamma$, which are based on the
kinematic suppression due to the larger mass of the higher
resonances given by the factor $(1-m^2_\Lambda/m^2_{\Lambda_b})^3$
\cite{Hiller-Kagan}. They do not take into account differences in
the form factors, nor a possible spin-dependence
of the decay probability, both to be determined by experiment.
Judging from recent data on $B\rightarrow K^*\gamma$ decays
\cite{HFAG} and from dedicated form factor studies on semi-leptonic
$B$ and $B_{s}$ meson decays \cite{Veseli-Olsson}, we may expect
these estimates to be correct up to a factor of 2--3 only. To
evaluate the $\Lambda(X)\rightarrow pK$ decay probabilities we use
the rough $\mathcal{B}(\Lambda(X)\rightarrow N\overline{K})$
estimates given in the Table and assume equal probabilities for
decays to $pK^-$ and $n\overline{K}\mbox{}^0$ from isospin coupling,
thereby neglecting possible suppression effects from angular
momentum barriers.

\begin{table}[th]
    \begin{center}
        \caption{Table of $\Lambda$ resonances decaying to $pK$ that are established
         with at least a fair degree of certainty.
         The listed widths $\Gamma$ and branching fractions $\mathcal{B}_\mathrm{tot}$ are those
         used to produce Fig.~\ref{fig:LbpKspec}.
         In the Table,
     $\mathcal{B}_{N\overline{K}} \equiv \mathcal{B}(\Lambda(X)\rightarrow N\overline{K})$,
         $\mathcal{B}_{\Lambda(X)\gamma} \equiv \mathcal{B}(\Lambda_b\rightarrow \Lambda(X)\gamma)$, and
     $\mathcal{B}_\mathrm{tot} \equiv \mathcal{B}(\Lambda_b\rightarrow \Lambda(X)\gamma\rightarrow pK\gamma$).
     The values for $\mathcal{B}_{N\overline{K}}$ are estimates based on data
         compiled in Ref.~\cite{PDG04}, whereas the $\mathcal{B}_{\Lambda(X)\gamma}$ are
         our estimates derived from simple kinematic suppression (see text).}
         \label{tab:lambdares}
        \begin{tabular*}{\columnwidth}{@{\extracolsep{\fill}}llrrrr}
        \hline
        $\Lambda(X)$ &
        $L_{I\cdot 2J}$ &
        $\Gamma$ &
        $\mathcal{B}_{N\overline{K}}$  &
        $\mathcal{B}_{\Lambda(X)\gamma}$ &
        $\mathcal{B}_\mathrm{tot}$\\
        &
        &
        (MeV) &
        (\%) &
        ($10^{-5}$) &
        ($10^{-5}$)\\
        \hline
        $\Lambda(1520)$ & $D_{03}$ & 15.6 &  45 & 5.84 & 1.31 \\
        $\Lambda(1600)$ & $P_{01}$ & 150  &  22 & 5.69 & 0.65 \\
        $\Lambda(1670)$ & $S_{01}$ &  35  &  25 & 5.56 & 0.69 \\
        $\Lambda(1690)$ & $D_{03}$ &  60  &  25 & 5.52 & 0.69 \\
        $\Lambda(1800)$ & $S_{01}$ & 300  &  32 & 5.30 & 0.84 \\
        $\Lambda(1810)$ & $P_{01}$ & 150  &  35 & 5.28 & 0.92 \\
        $\Lambda(1820)$ & $F_{05}$ &  80  &  60 & 5.26 & 1.57 \\
        $\Lambda(1830)$ & $D_{05}$ &  95  &   6 & 5.24 & 0.15 \\
        $\Lambda(1890)$ & $P_{03}$ & 100  &  22 & 5.12 & 0.56 \\
        $\Lambda(2100)$ & $G_{07}$ & 200  &  30 & 4.67 & 0.70 \\
        $\Lambda(2110)$ & $F_{05}$ & 200  &  15 & 4.65 & 0.34 \\
        $\Lambda(2350)$ & $H_{09}$ & 150  &  12 & 4.12 & 0.28 \\
        \hline
       \end{tabular*}
    \end{center}
\end{table}

\begin{figure}[bh]
\hskip -1cm
      \includegraphics[width=16cm]{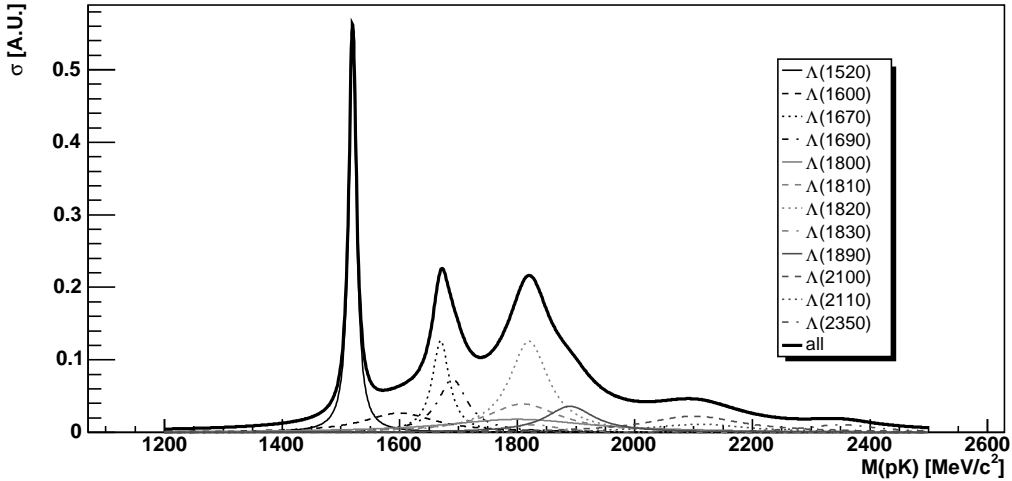}
\vskip -1 cm
    \caption{Approximate $pK$ effective (``invariant'') mass spectrum from
    $\Lambda_b\rightarrow pK\gamma$
    decays, as obtained with the values in Table~\ref{tab:lambdares} by adding up
    simple non-relativistic Breit-Wigner forms.
    A possible non-resonant contribution is neglected, as are interference effects
    between the various resonances.}
    \label{fig:LbpKspec}
\end{figure}

Figure~\ref{fig:LbpKspec} illustrates the $pK$ effective mass
spectrum resulting from our simplifying assumptions. While the true
spectrum, to be measured experimentally, may look different in
detail, it is still useful to have a general overview of the
$\Lambda(X)$ resonance properties, which allows us to identify the
most promising decay modes. The mass spectrum is likely to
feature the three rather distinct peaks visible in
Fig.~\ref{fig:LbpKspec}. The first and most prominent of these peaks
is due to the well-established $\Lambda(1520)$. Since this resonance
has spin 3/2, the extraction of $\alpha_\gamma$, and thus $r$, via
angular decay distributions is not straight-forward. We will see in
Sec.~\ref{subsec:J=3/2} that it is possible under certain
conditions. The second peak is made up of the $\Lambda(1670)$ (spin
1/2) and $\Lambda(1690)$ (spin 3/2) resonances. It may be assumed
that the different angular decay distributions allow for the
disentanglement of the two resonances, so that $\alpha_\gamma$ can
be extracted from a combined fit applied to events in that region. A
possible third peak is probably dominated by the $\Lambda(1820)$.
Since this resonance has spin 5/2, we do not consider it useful for
the determination of the photon polarization.

\section{Decay angular distributions for radiative $\boldsymbol{\Lambda_b}$ decays}

In the helicity formalism \cite{Jacob-Wick-Richman}, we may write
down the decay amplitude for the general case of the decay
$\Lambda_b\rightarrow \Lambda\gamma\rightarrow ph\gamma$ ($h =
K,\pi$) as
\begin{equation}\label{eq:A}
    A = \sum_{\lambda_\Lambda} D^{J_\Lambda *}_{\lambda_\Lambda,\lambda_p}
                                                 (\phi_p,\theta_p,-\phi_p)
        D^{J*}_{M,\lambda_\Lambda-\lambda_\gamma}
        (\phi_\Lambda,\theta_\Lambda,-\phi_\Lambda)
        C_{\lambda_\Lambda,\lambda_\gamma}
        E_{\lambda_p},
\end{equation}
where $\lambda_i$ ($J_i$) is the helicity (spin) of particle $i$,
$J$ and $M$ refer to the $\Lambda_b$ spin and its projection along the
(arbitrary) quantization axis, respectively;
the polar and azimuthal angles $\theta_\Lambda$ and $\phi_\Lambda$, defined
in the $\Lambda_b$ rest frame, give the direction of the $\Lambda$ momentum
relative to the quantization axis;
the angles $\theta_p$ and $\phi_p$, defined in the $\Lambda$ rest frame,
give the direction of the proton momentum relative to the $\Lambda$ flight direction.
The quantities $C$ and $E$ parameterize the intrinsic helicity amplitudes
for the decays $\Lambda_b\rightarrow \Lambda\gamma$ and $\Lambda\rightarrow ph$,
respectively.
If parity is conserved (i.e., in strong decays to $pK$),
$|E_{\lambda_p}| = |E_{-\lambda_p}|$.

The decay probability is obtained by squaring the amplitude and
summing over the final state helicities, which are not measured by
the experiment,
\begin{equation}\label{eq:w}
 w = \sum_{M, \lambda_\gamma, \lambda_p} \rho_{MM}|A|^2,
\end{equation}
where the polarization density matrix $\rho$ takes account of the $\Lambda_b$
polarization with respect to the quantization axis.
Since we do not consider correlations between the production and decay mechanisms,
the non-diagonal elements of $\rho$ average out to zero \cite{Ueda-Okubo}
whereas the diagonal elements are characterized by
$\rho_{\frac{1}{2},\frac{1}{2}}+\rho_{-\frac{1}{2},-\frac{1}{2}} =
\operatorname{Tr}\rho = 1$ and the $\Lambda_b$ polarization
$P_{\Lambda_b} = \rho_{\frac{1}{2},\frac{1}{2}}-\rho_{-\frac{1}{2},-\frac{1}{2}}$.
The explicit form of the decay probability $w$ then depends on the spin of the
intermediate $\Lambda$.
We separately treat the cases $J_\Lambda = 1/2$ and $J_\Lambda = 3/2$.


\subsection{The case $J_\Lambda = \frac{1}{2}$ }
\label{subsec:J=1/2}

Angular distributions for the spin-1/2 case have been given in
Ref.~\cite{Hiller-Kagan}. We re-derive them here as a warm-up, and
to introduce our notation. For $J_\Lambda = 1/2$ we have only two
allowed helicity combinations which we may identify by the total
helicity $\lambda \equiv \lambda_\Lambda - \lambda_\gamma = \pm1/2$,
with corresponding amplitudes $C_\lambda$. The decay amplitude
(\ref{eq:A}) becomes
\begin{equation}
    A = \sum_{\lambda} D^{\frac{1}{2} *}_{\lambda_\Lambda,\lambda_p}(\phi_p,\theta_p,-\phi_p)
        D^{\frac{1}{2}*}_{M,\lambda}(\phi_\Lambda,\theta_\Lambda,-\phi_\Lambda)
        C_\lambda E_{\lambda_p},
\end{equation}
therefore (dropping the argument angles for better readability)
\begin{equation*}
    |A|^2 =
        \sum_{\lambda} |D^{\frac{1}{2}}_{\lambda_\Lambda,\lambda_p}|^2
        |D^{\frac{1}{2}}_{M,\lambda}|^2 |C_\lambda|^2 |E_{\lambda_p}|^2,
\end{equation*}
and the decay probability (\ref{eq:w}) becomes
\begin{equation*}
 w_\frac{1}{2} =
       \sum_{\lambda_p,\lambda}|C_\lambda|^2|E_{\lambda_p}|^2| D^{\frac{1}{2}}_{\lambda_{\Lambda},\lambda_p}|^2
       \left[\rho_{\frac{1}{2},\frac{1}{2}} |D^\frac{1}{2}_{\frac{1}{2},\lambda}|^2
            +\rho_{-\frac{1}{2},-\frac{1}{2}} |D^\frac{1}{2}_{-\frac{1}{2},\lambda}|^2\right].
\end{equation*}
Inserting
$   D^j_{m,m'}(\alpha, \beta, \gamma) = e^{i\alpha m'}
    d^j_{m,m'}(\beta)e^{-i\gamma m} $
and explicit expressions for the $d$-functions it is a matter of straight-forward
algebra to obtain
\begin{equation}\label{eq:w1/2}
    w_\frac{1}{2} \propto
    1-\alpha_{p, \frac{1}{2}}P_{\Lambda_b}\cos\theta_p\cos\theta_\Lambda -
    \alpha_{\gamma, \frac{1}{2}}(\alpha_{p, \frac{1}{2}}\cos\theta_p -
    P_{\Lambda_b}\cos\theta_\Lambda),
\end{equation}
where we have defined, in addition to the $\Lambda_b$ polarization $P_{\Lambda_b}$,
the photon asymmetry
\begin{equation} \label{eq:asy_gamma}
  \alpha_{\gamma, \frac{1}{2}}
  =\frac{|C_\frac{1}{2}|^2-|C_{-\frac{1}{2}}|^2}{|C_\frac{1}{2}|^2+|C_{-\frac{1}{2}}|^2}
\end{equation}
and the proton asymmetry
\begin{equation}
  \alpha_{p, \frac{1}{2}}
  =\frac{|E_\frac{1}{2}|^2-|E_{-\frac{1}{2}}|^2}{|E_\frac{1}{2}|^2+|E_{-\frac{1}{2}}|^2}.
\end{equation}
Clearly, $\alpha_{p,\frac{1}{2}} = 0$ for $\Lambda(X)\rightarrow pK$ due to parity conservation,
but $\alpha_{p,\frac{1}{2}} = 0.642\pm0.013$ for $\Lambda(1116)\rightarrow p\pi$ \cite{PDG04}.
Integration over the solid angle elements $d\Omega_p$ and $d\Omega_\gamma$ finally
yields the well-known angular distributions
\begin{equation}\label{eq:ang_gamma}
    \frac{d\Gamma}{d\cos\theta_\gamma} \propto 1-\alpha_{\gamma, \frac{1}{2}}
     P_{\Lambda_b}\cos\theta_\gamma ,
\end{equation}
where $\cos\theta_\gamma = -\cos\theta_\Lambda$, and
\begin{equation}\label{eq:ang_p}
    \frac{d\Gamma}{d\cos\theta_p} \propto 1-\alpha_{\gamma, \frac{1}{2}}
     \alpha_{p, \frac{1}{2}}\cos\theta_p .
\end{equation}

These distributions show that, if the $\Lambda_b$ polarization is known,
$\Lambda_b\rightarrow\Lambda(1116)\gamma$ decays afford two independent ways to assess
the polarization of the emitted photon, as pointed out in Ref.~\cite{Hiller-Kagan},
whereas for decays to spin-1/2 $\Lambda$ resonances only the photon
distribution is useful in that respect.
Indeed, comparing (\ref{eq:asy_gamma}) with (\ref{eq:alpha}) we identify
\begin{equation}
  \alpha_\gamma = \alpha_{\gamma,\frac{1}{2}},
\end{equation}
i.e., $r$ can be extracted directly from the angular distributions.
We note that in a real experiment, possible selection bias effects in the
photon angular distribution can easily be corrected for by applying the
same selection to the abundant channel $B\rightarrow K^*\gamma \rightarrow
K\pi\gamma$, which has a very similar topology but no intrinsic photon
asymmetry.


\subsection{The case $J_\Lambda = \frac{3}{2}$ }
\label{subsec:J=3/2}

In the spin-3/2 case, the number of allowed helicity combinations increases
to four, defined by $\lambda_\Lambda = \pm 3/2$ (with $\lambda_\gamma = \pm 1$,
$\lambda = \pm 1/2$) and $\lambda_\Lambda = \pm 1/2$ ($\lambda_\gamma = \pm 1$,
$\lambda = \mp 1/2$) and governed by corresponding helicity amplitudes
$C_{\lambda_\Lambda,\lambda_\gamma}$.
In this case the decay amplitude (\ref{eq:A}) becomes:
\begin{equation}
    A = \sum_{\lambda_\Lambda} D^{\frac{3}{2} *}_{\lambda_\Lambda,\lambda_p}(\phi_p,\theta_p,-\phi_p)
        D^{\frac{1}{2}*}_{M,\lambda_\Lambda-\lambda_\gamma}(\phi_\Lambda,\theta_\Lambda,-\phi_\Lambda)
        C_{\lambda_\Lambda,\lambda_\gamma}
        E_{\lambda_p}
\end{equation}
Proceeding analogously to the spin-1/2 case we obtain for the squared amplitude
\begin{align*}
    |A|^2 \propto &
     \sum_{\lambda_\Lambda,\lambda'_\Lambda}
     d^{\frac{3}{2}}_{\lambda_\Lambda,\lambda_p} (\theta_p)
     d^{\frac{3}{2}}_{\lambda'_\Lambda,\lambda_p}(\theta_p)
     d^{\frac{1}{2}}_{M,\lambda_\Lambda-\lambda_\gamma}(\theta_\Lambda)
     d^{\frac{1}{2}}_{M,\lambda'_\Lambda-\lambda_\gamma}(\theta_\Lambda) \\
     & \times  e^{i(\phi_\Lambda+\phi_p)(\lambda'_\Lambda-\lambda_\Lambda)}
     C_{\lambda_\Lambda,\lambda_\gamma}C^*_{\lambda'_\Lambda,\lambda_\gamma} \\
     = &
    \underbrace{|C_{\frac{3}{2},1}|^2|d^{\frac{3}{2}}_{\frac{3}{2},\lambda_p}(\theta_p)
      d^{\frac{1}{2}}_{M,\frac{1}{2}}(\theta_\Lambda)|^2}_{|A_1|^2} +
    \underbrace{|C_{-\frac{3}{2},-1}|^2|
      d^{\frac{3}{2}}_{-\frac{3}{2},\lambda_p}(\theta_p)
      d^{\frac{1}{2}}_{M,-\frac{1}{2}}(\theta_\Lambda)|^2 }_{|A_2|^2} \\
    & +\underbrace{|C_{\frac{1}{2},1}|^2|
      d^{\frac{3}{2}}_{\frac{1}{2},\lambda_p}(\theta_p)
      d^{\frac{1}{2}}_{M,-\frac{1}{2}}(\theta_\Lambda)|^2}_{|A_3|^2}
      +\underbrace{|C_{-\frac{1}{2},-1}|^2|
      d^{\frac{3}{2}}_{-\frac{1}{2},\lambda_p}(\theta_p)
      d^{\frac{1}{2}}_{M,\frac{1}{2}}(\theta_\Lambda)|^2}_{|A_4|^2} \\
    & +\underbrace{2~\operatorname{Re}
      \{C^*_{\frac{3}{2},1}C_{\frac{1}{2},1}e^{i(\phi_\Lambda+\phi_p)}\}
      d^{\frac{3}{2}}_{\frac{3}{2},\lambda_p}(\theta_p)
      d^{\frac{1}{2}}_{M,\frac{1}{2}}(\theta_\Lambda)
      d^{\frac{3}{2}}_{\frac{1}{2},\lambda_p}(\theta_p)
      d^{\frac{1}{2}}_{M,-\frac{1}{2}}(\theta_\Lambda)}_{|A_5|^2} \\
    & +\underbrace{2~\operatorname{Re}
      \{C^*_{-\frac{3}{2},-1}C_{-\frac{1}{2},-1}e^{-i(\phi_\Lambda+\phi_p)}\}
      d^{\frac{3}{2}}_{-\frac{3}{2},\lambda_p}(\theta_p)
      d^{\frac{1}{2}}_{M,-\frac{1}{2}}(\theta_\Lambda)
      d^{\frac{3}{2}}_{-\frac{1}{2},\lambda_p}(\theta_p)
      d^{\frac{1}{2}}_{M,\frac{1}{2}}(\theta_\Lambda)}_{|A_6|^2} ,
\end{align*}
where we have dropped the overall factor $|E_{\lambda_p}|^2$, which is constant
due to parity conservation.
For the decay probability (\ref{eq:w}) we then have
\begin{equation}\label{eq:w3/2_app}
w_{\frac{3}{2}} = \sum_{i=1}^6 \sum_{M,\lambda_\Lambda,\lambda_p} \rho_{MM} |A_i|^2
=\sum_{i=1}^6 w_i
\end{equation}
with
\begin{alignat*}{4}
 w_1 & =
    \frac{3}{8}  ~ &&  |C_{\frac{3}{2},1}|^2 ~ &&
    \sin^2\theta_p ~ && (1+P_{\Lambda_b} \cos\theta_\Lambda) ; \\
 w_2 & =
    \frac{3}{8} ~ &&  |C_{-\frac{3}{2},-1}|^2 ~ &&
    \sin^2\theta_p ~ && (1-P_{\Lambda_b} \cos\theta_\Lambda) ; \\
 w_3 & =
    \frac{1}{8} ~ &&  |C_{\frac{1}{2},1}|^2 ~ &&
    (3\cos^2\theta_p+1) ~ &&
    (1-P_{\Lambda_b} \cos\theta_\Lambda) ; \\
 w_4 & =
    \frac{1}{8} ~ && |C_{-\frac{1}{2},-1}|^2 ~ &&
    (3\cos^2\theta_p+1) ~ &&
    (1+P_{\Lambda_b} \cos\theta_\Lambda) ;
\end{alignat*}
\begin{alignat*}{3}
 w_5  = &
    \frac{\sqrt{3}}{2} ~ &&
    \operatorname{Re}\{C^*_{\frac{3}{2},1}C_{\frac{1}{2},1}
    e^{i(\phi_\Lambda+\phi_p)}\} &&
    ~  \cos\theta_p~\sin\theta_p~ \sin\theta_\Lambda ;  \\
 w_6  = &
    \frac{\sqrt{3}}{2} ~ &&
    \operatorname{Re}\{C^*_{-\frac{3}{2},1}C_{-\frac{1}{2},1}
    e^{-i(\phi_\Lambda+\phi_p)}\} &&
    ~  \cos\theta_p~\sin\theta_p~ \sin\theta_\Lambda .
\end{alignat*}
Again we integrate over the appropriate solid angle elements and
get
\begin{equation}\label{eq:ang_gamma_3/2}
    \frac{d\Gamma}{d\cos\theta_\gamma} \propto
    1-\alpha_{\gamma, \frac{3}{2}} P_{\Lambda_b} \cos\theta_\gamma ,
\end{equation}
where now the photon asymmetry parameter is defined as
\begin{equation}\label{eq:alpha_3/2}
\alpha_{\gamma, \frac{3}{2}} =
\frac{\overbrace{|C_{\frac{3}{2},1}|^2+
     |C_{-\frac{1}{2},-1}|^2}^{\lambda_\Lambda-\lambda_\gamma=1/2}-
      \overbrace{|C_{-\frac{3}{2},-1}|^2-
     |C_{\frac{1}{2},1}|^2}^{\lambda_\Lambda-\lambda_\gamma=-1/2}}
     {|C_{\frac{3}{2},1}|^2+|C_{-\frac{1}{2},-1}|^2+
      |C_{-\frac{3}{2},-1}|^2+|C_{\frac{1}{2},1}|^2} ,
\end{equation}
i.e., it describes the asymmetry of the $\Lambda_b$ spin projection
with respect to the photon momentum.
It is obvious from (\ref{eq:alpha_3/2}) that the extraction of $\alpha_\gamma$,
and therefore the photon polarization in the fundamental $b\rightarrow s\gamma$
process, from $\alpha_{\gamma,\frac{3}{2}}$ is only possible if we know the
relative strengths of the $m=1/2$ and $m=3/2$ amplitudes.
The parameter $\eta$, defined as
\begin{equation}\label{eq:eta}
\eta =
\frac{|C_{\frac{3}{2},1}|^2}{|C_{\frac{1}{2},1}|^2} =
\frac{|C_{-\frac{3}{2},-1}|^2}{|C_{-\frac{1}{2},-1}|^2}
\end{equation}
(where the second equals sign is justified by parity conservation in the
ha\-dron\-i\-za\-tion process), allows us to relate $\alpha_{\gamma,\frac{3}{2}}$
and $\alpha_{\gamma}$ in a simple way:
\begin{equation}\label{eq:alpha_gamma_3/2}
\alpha_{\gamma, \frac{3}{2}} =\frac{1-\eta}{1+\eta}~\alpha_{\gamma}
\end{equation}
We can determine $\eta$ experimentally from the proton angular distribution.
Indeed, integration of Eq.~(\ref{eq:w3/2_app}) over $\theta_\Lambda$ yields
\begin{equation}\label{eq:ang_p_3/2}
    \frac{d\Gamma}{d\cos\theta_p} \propto
    1 - \alpha_{p,\frac{3}{2}} \cos^2\theta_p ,
\end{equation}
with the proton asymmetry parameter
\begin{equation}\label{eq:alpha_p_3/2}
\alpha_{p,\frac{3}{2}} =
\frac{\overbrace{|C_{\frac{3}{2},1}|^2+|C_{-\frac{3}{2},-1}|^2}^{|\lambda_\Lambda|=3/2}-
\overbrace{|C_{\frac{1}{2},1}|^2-|C_{-\frac{1}{2},-1}|^2}^{|\lambda_\Lambda|=1/2}}
{|C_{\frac{3}{2},1}|^2+|C_{-\frac{3}{2},-1}|^2+\frac{1}{3}(|C_{-\frac{1}{2},-1}|^2+|C_{\frac{1}{2},1}|^2)}
= \frac{\eta-1}{\eta+\frac{1}{3}} .
\end{equation}
The proton polar angle distribution is symmetric around $\cos\theta_p = 0$, as expected
for a strong decay, but it still allows us to extract a value for $\eta$.
The determination of $\alpha_\gamma$ from a combined measurement of photon
and proton angular distributions is then possible according to
\begin{equation}\label{eq:alpha_gamma_1/2_3/2}
\alpha_{\gamma} = \frac{1}{2}~\alpha_{\gamma,\frac{3}{2}}
\left(1-\frac{3}{\alpha_{p, \frac{3}{2}}}\right),
\end{equation}
if $\eta$ is sufficiently far away from 1 (equal probability for
$m=1/2$ and $m=3/2$).
For $\eta \ll 1$ the $m=1/2$ amplitude dominates, $\alpha_{p,3/2} \simeq -3$,
and $\alpha_{\gamma, \frac{3}{2}} \simeq \alpha_{\gamma}$.
In the case where the $m=3/2$ amplitude dominates ($\eta \gg 1$)
$\alpha_{p,\frac{3}{2}} \simeq 1$ and
$\alpha_{\gamma, \frac{3}{2}} \simeq - \alpha_{\gamma}$.

\section{Experimental prospects for a photon polarization measurement
in $\boldsymbol{\Lambda_b\rightarrow pK\gamma}$}

We now compare the experimental prospects for a measurement of the
photon polarization (parameter $|r|$) in $\Lambda_b\rightarrow
\Lambda(X)\gamma\rightarrow pK\gamma$ decays to those in
$\Lambda_b\rightarrow \Lambda(1116)\gamma\rightarrow p\pi\gamma$
decays at a hadron collider. For the $\Lambda_b$ polarization we
will assume a mean value \cite{Ajaltouni} and experimental error
\cite{Hrivnac} of $P_{\Lambda_b} = 0.20\pm 0.01 $. For the sake of a
concrete estimate of the sensitivity in $|r|$ we fix the number of
fully reconstructed $\Lambda_b\rightarrow \Lambda(1520)\rightarrow
pK\gamma$ to $10^4$ and scale the event yields for the other
resonance channels according to their branching fractions. We note
that this number is arbitrary but realistic. Indeed, the LHCb
collaboration for example expects an annual yield of 35\,000 events
containing the topologically very similar decay $B\rightarrow
K^*\gamma\rightarrow K\pi\gamma$ \cite{LHCb-Reopt-TDR}. Factoring in
the relevant production rates and branching fractions, but assuming
equal reconstruction efficiencies, we find that it would take LHCb a
little more than three years to collect $10^4$ $\Lambda_b\rightarrow
\Lambda(1520)\rightarrow pK\gamma$ decays.

Clearly, the total reconstruction efficiency, including trigger, for
$\Lambda_b\rightarrow$ $\Lambda(1116)\gamma$ $\rightarrow p\pi\gamma$ decays will be
significantly lower than that for $\Lambda_b\rightarrow \Lambda(X)\gamma\rightarrow
pK\gamma$ decays.
Since it is hard to predict the experimental difficulties at this time,
we not only consider a {\em default} scenario where the reconstruction efficiency is
ten times worse with respect to $\Lambda_b\rightarrow
\Lambda(1520)\gamma\rightarrow pK\gamma$, but also a {\em best} (very optimistic)
and a {\em worst} (very pessimistic) scenario in which the reconstruction
efficiency is assumed to be equal and a hundred times worse, respectively.

\begin{figure}[p]
  \includegraphics[width=12cm]{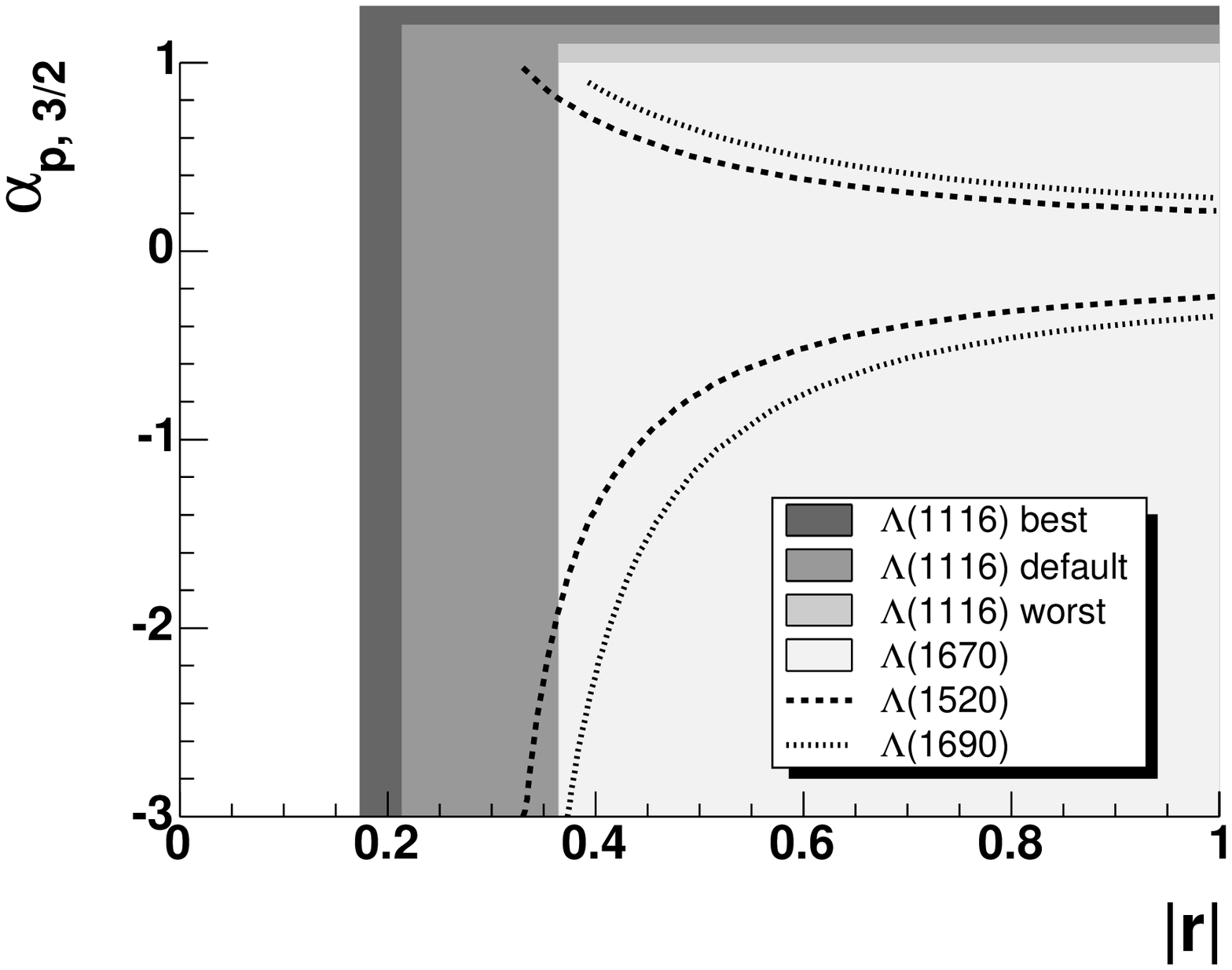}
  \includegraphics[width=12cm]{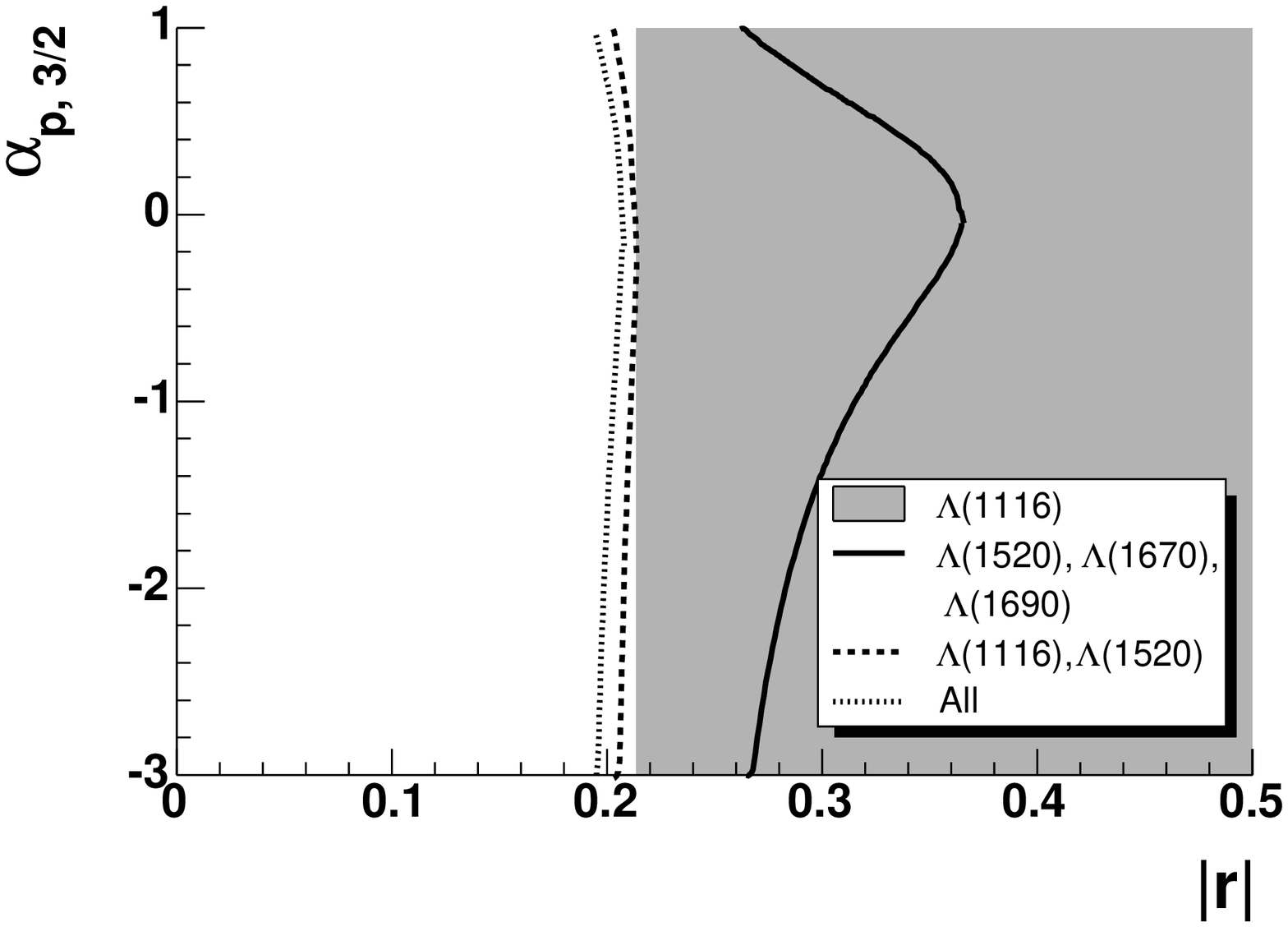}
  \caption{Experimental reach for $|r|$ (as a function of $\alpha_{p,\frac{3}{2}}$ for decays involving
    a spin-3/2 $\Lambda(X)$).
    The plots show the values of $|r|$ that can be probed at
    3$\sigma$ (standard deviation) significance
    in single (top) and combined (bottom) measurements, at a hadron collider experiment
    capable of collecting $10^4$
    $\Lambda_b\rightarrow \Lambda(1520)\gamma\rightarrow pK\gamma$ decays.
    The ranges are to be read from left to right starting from the curves.
    The three ranges for the $\Lambda(1116)$ in the top plot correspond to the best,
    default and worst reconstruction scenarios (see text).
    The bottom plot is based on the default reconstruction scenario and on the assumption
    of equal $\alpha_{p,3/2}$ for $\Lambda(1520)$ and $\Lambda(1690)$.}
    \label{fig:r_sign}
\end{figure}

In Fig.~\ref{fig:r_sign} we show the expected experimental (statistical only)
reach for the parameter $|r|$ at a hadron collider,
as obtained under the above assumptions and with the error evaluation described
in Appendix~\ref{app:err}.
In the top plot the reach is shown
separately for the three resonances $\Lambda(1520)$, $\Lambda(1670)$, and
$\Lambda(1690)$.
The sensitivity curves are compared with the reach obtained using the $\Lambda(1116)$,
where both the photon and the proton asymmetry contribute to the measurement,
in the three scenarios of different reconstruction efficiency.
Note that for the spin-3/2 $\Lambda$ resonances, the expected reach is a function
of the parameter $\alpha_{p,\frac{3}{2}}$, to be determined by the experiment.
The bottom plot in Fig.~\ref{fig:r_sign} illustrates the experimental reach
for various plausible combinations of measurements in the $\Lambda(1116)$ default
scenario:
\begin{itemize}
\item
the case where only $\Lambda(1520)$, $\Lambda(1670)$ and $\Lambda(1690)$ are
available ($\Lambda(1116)$ cannot be reconstructed),
\item
the case where only $\Lambda(1116)$ and $\Lambda(1520)$ contribute
($\Lambda(1670)$ and $\Lambda(1690)$ cannot be disentangled), and
\item
the case where all four $\Lambda$ states enter the determination of $|r|$.
\end{itemize}

We see that in the case of the $\Lambda(1116)$ the availability of
both the photon (\ref{eq:ang_gamma}) and the proton asymmetry
(\ref{eq:ang_p}) for determining $|r|$ largely compensates even for
large losses in statistics due to reconstruction problems. Under our
default assumptions, decays to $\Lambda(1116)$ will allow a typical
hadron collider experiment to probe $|r|$ down to 0.21, whereas the
decays to $\Lambda$ resonances can only give constraints to about
0.33. A combination of measurements of the three $\Lambda$
resonances $\Lambda(1520)$, $\Lambda(1670)$ and $\Lambda(1690)$ can
probe down to 0.27 in the most favourable case.

\vskip 1cm

\begin{figure}[ht]
  \begin{center}
      \includegraphics[width=15 cm]{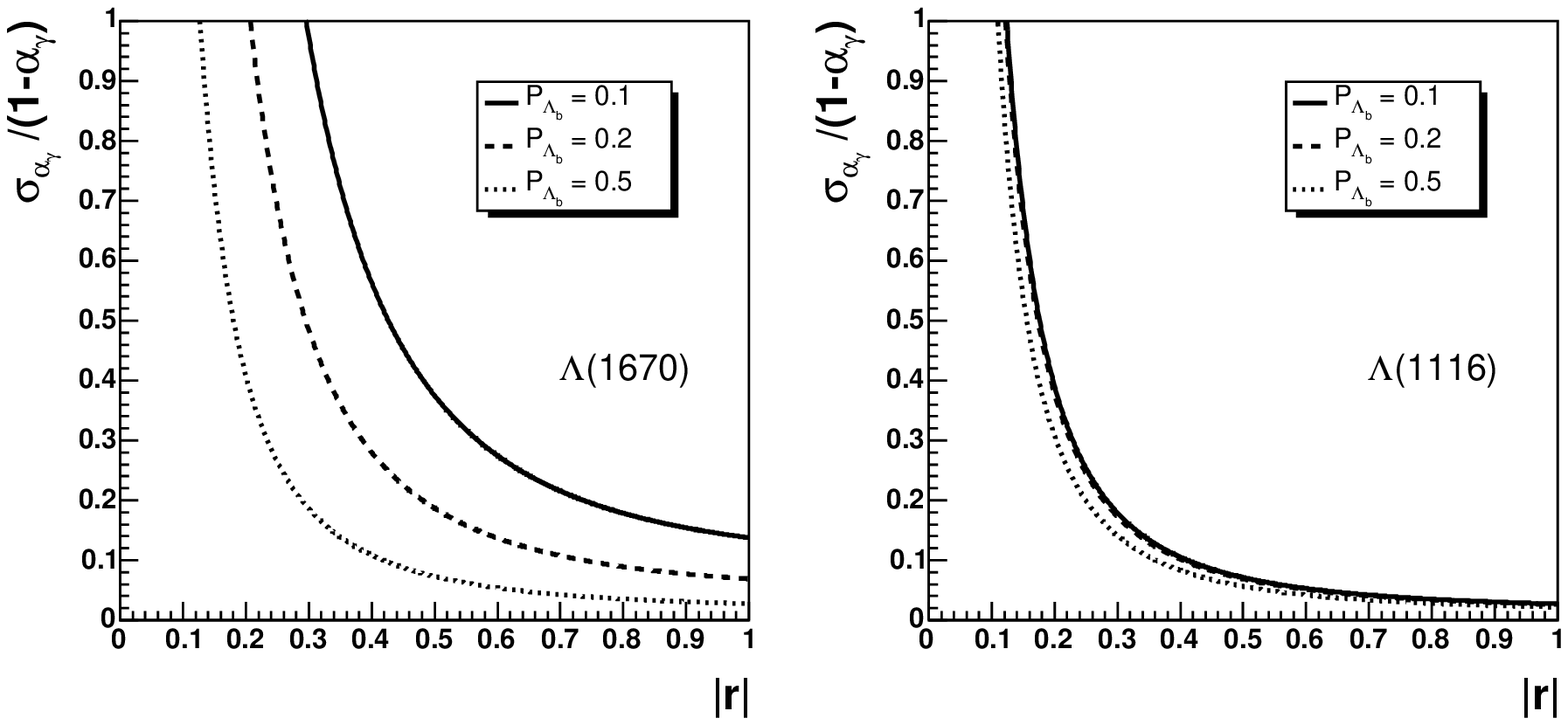}
  \vskip -.5 cm
    \caption{Expected relative statistical error on $(1-\alpha_\gamma)$ 
    as a function of $|r|$
    at a hadron collider experiment for the decays
    $\Lambda_b\rightarrow \Lambda(1670)\gamma$ (left) and
    $\Lambda_b\rightarrow \Lambda(1116)\gamma$ (right).
    The three curves represent different assumptions
    for the $\Lambda_b$ polarization $P_{\Lambda_b}$: 0.1 (solid),
    0.2 (dashed) and 0.5 (dotted).
    For $\Lambda_b \rightarrow \Lambda(1115) \gamma$ the default reconstruction
    scenario is assumed (see text).
    Event yields are as in Fig.~\ref{fig:r_sign}.}
  \label{fig:r_sign_pol}
  \end{center}
\end{figure}

Since our current estimate for the $\Lambda_b$ polarization at a
hadron collider may well be off by a large factor, it is worth
examining the dependence of the relative error of $|r|$ on
$P_{\Lambda_b}$. To illustrate the effect of the $\Lambda_b$
polarization on the statistical reach in $|r|$, we show in
Fig.~\ref{fig:r_sign_pol} for
$\Lambda_b\rightarrow\Lambda(1670)\gamma$ and
$\Lambda_b\rightarrow\Lambda(1116)\gamma$ (default reconstruction
scenario) the relative statistical error on $(1-\alpha_\gamma)$ 
(cf.\ Appendix~\ref{app:err}) as a function of
$|r|$ for the three cases $P_{\Lambda_b} = 0.1$, 0.2 and 0.5. As
expected, in the case of the decay to a $\Lambda$ resonance the
measurement of $|r|$ is much less robust against small values of the
$\Lambda_b$ polarization than in the case of the $\Lambda(1116)$,
where the proton asymmetry allows for a measurement of $|r|$ even if
the $\Lambda_b$ is not polarized at all.

\begin{figure}[ht]
  \begin{center}
      \includegraphics[width=15 cm]{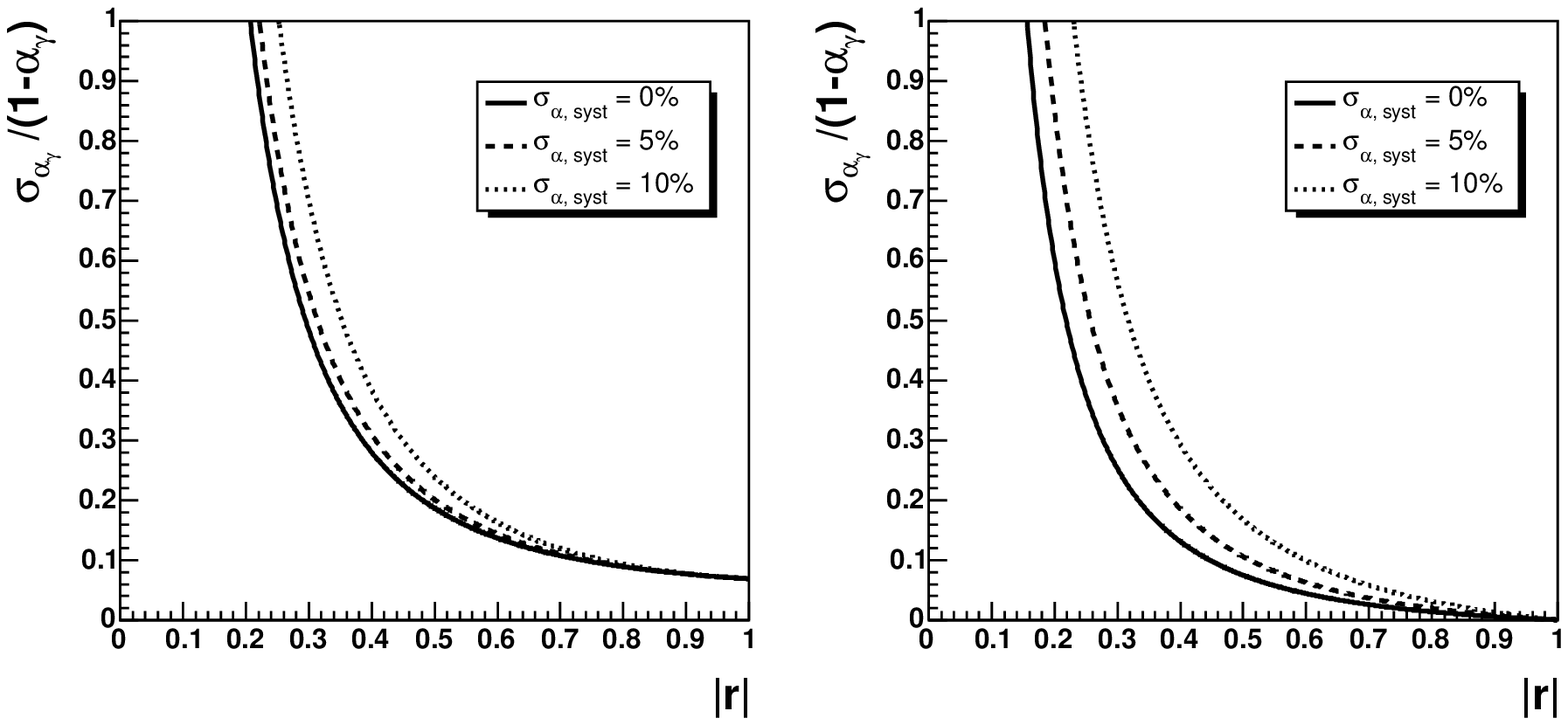}
  \vskip -.5 cm
    \caption{Expected relative total error on $(1-\alpha_\gamma)$
     as a function of $|r|$
    at a hadron collider experiment for the decay
    $\Lambda_b\rightarrow \Lambda(1670)\gamma$.
    The three curves represent different assumptions
    for the detector-related systematic relative error 
    $\sigma_{\alpha,\mathrm{syst}}$ on the measurement
    of the photon asymmetry $\alpha_\gamma$: 0\% (solid),
    5\% (dashed), and 10\% (dotted).
    Event yields are as in Fig.~\ref{fig:r_sign} (left plot), and 
    corresponding to infinite statistics (right plot).}
  \label{fig:r_syst}
  \end{center}
\vspace{2cm}
\end{figure}

Another concern arises from systematic errors in the 
measurement of the photon and proton asymmetries. 
Similar to the reconstruction efficiency, these uncertainties
depend on the specific experimental setup and cannot be estimated
in a general way.
Nevertheless, given the cancellation of a large class of experimental
effects in asymmetry measurements, we may assume that these errors 
will not exceed the few-percent level for the parameter $\alpha_\gamma$.
For the sake of illustration, we plot in Fig.~\ref{fig:r_syst}
the expected total relative error on $(1-\alpha_\gamma)$
as a function of $|r|$
in the presence of a systematic error on $\alpha_\gamma$ of 
0\%, 5\%, and 10\% for the $\Lambda(1670)$ example.
To explore the ultimate sensitivity we show the same curves for 
infinite statistics. 
We see that even in the case of vanishing statistical and (internal) 
systematic errors, the sensitivity would still be limited to about 
$|r|>0.25$.
In the case of the $\Lambda(1116)$ we find a sensitivity limit of about 
0.15.
These limits are a consequence of our assumptions on the uncertainties
of the $\Lambda_b$ production polarization and the $\Lambda$ weak decay
parameter.

\vskip 1cm

It is interesting to compare the experimental reach for $|r|$ at a
hadron collider with prospects at the $B$ factories. The only method
applied so far at the $B$ factories is the one relying on
$B$-$\overline{B}$ interference \cite{Atwood-Gronau-Soni}, where the
CP-violation parameter $\mathcal{S}$ \cite{Kirkby-Nir} has been
measured for the decay $B^0 \rightarrow K^0_S \pi^0 \gamma$. The
amplitude ratio $|r|$ may be extracted from $\mathcal{S}_{K^0_S \pi^0
\gamma}$ according to $\mathcal{S}_{K^0_S \pi^0 \gamma} =
2|r|\sin2\beta$ \cite{Atwood-Gronau-Soni}, where $\sin2\beta$ is the
by now well known CKM parameter describing the $B$-$\overline{B}$
mixing phase. (Note that this method has a linear sensitivity in $|r|$
as opposed to the $|r|^2$ dependence of $\alpha_\gamma$, the principal
observable in radiative $\Lambda_b$ decays). The most recent
measurement of $\mathcal{S}_{K^0_S \pi^0 \gamma}$ has been presented
by the Belle Collaboration \cite{Ushiroda}. Based on 5.35 $\times
10^8$ $B\overline{B}$ pairs, Belle finds $\mathcal{S}_{K^0_S \pi^0
\gamma} = -0.10 \pm0.31 \pm0.07$, where the first error is
statistical and the second is systematic. Using the latest world
average $\sin2\beta = 0.674 \pm0.026$ \cite{HazumiICHEP06}, the
error on $\mathcal{S}$ translates to an error on $|r|$ of $\sigma_{|r|} =
0.23$; in other words: $|r|$ would have to be as large as 0.7 for the
observation of a right-handed component at the 3$\sigma$ level at
today's $B$ factories. Recent assessments of the physics potential
of a next-generation $B$ factory find uncertainties on
$\mathcal{S}_{K^0_S \pi^0 \gamma}$ of 0.1 and 0.03 for integrated
luminosities of 5 and 50 ab$^{-1}$, respectively
\cite{HazumiFlavorLHC06}. This would correspond to 3$\sigma$ reaches
of $|r|>0.22$ and $|r|>0.07$, respectively, which may in principle be 
compared to the curves in Fig.~\ref{fig:r_sign}.
One should, however, keep in mind that the curves in Fig.~\ref{fig:r_sign}
do not contain effects from detector-related systematic errors, which could 
substantially limit the sensitivity at hadron colliders 
(Fig.~\ref{fig:r_syst}).

\section{Conclusion}

To summarize, radiative $\Lambda_b$ decays to $\Lambda$ resonances above the nucleon-kaon
threshold provide an interesting alternative to the experimentally challenging decay
$\Lambda_b\rightarrow \Lambda(1116)\gamma$ for assessing the photon polarization in the
quark transition $b\rightarrow s\gamma$ at hadron colliders.
The principal unknown for spin-3/2 resonances such as $\Lambda(1520)$ and $\Lambda(1690)$,
the repartition of $m=1/2$ and $m=3/2$ amplitudes, can be extracted directly from
experiment.

We have studied the experimental prospects for constraining the presence of anomalous
right-handed currents in the $b\rightarrow s\gamma$ transition, parameterized by the
ratio of Wilson coefficients $r = C'_7/C_7$, in the context of a generic hadron collider
experiment.
Our comparison between the decays $\Lambda_b\rightarrow \Lambda(1116)\gamma\rightarrow
p\pi\gamma$ and $\Lambda_b\rightarrow \Lambda(X)\gamma\rightarrow pK\gamma$ shows
that although the decay to $\Lambda(1116)$ offers by far the best sensitivity thanks
to the simultaneous contributions from photon and proton asymmetries, a combined
analysis of decays to $\Lambda$ resonances is capable of recovering a large
fraction of the discovery range in $|r|$ in the case where the $\Lambda$ is not
detectable due to its escape of the inner detector system.

\appendix

\section{Statistical error estimation}
\label{app:err}

In the linear approximation, the relative statistical error on $|r|$ is given 
by
\begin{equation} \label{eq:err_r}
    \sigma_{|r|}=  
    \frac{1}{|r|}\frac{\sigma_{\alpha_{\gamma}}}{(1+\alpha_{\gamma})^2}.
\end{equation}
We note, however, that the application of this formula can give misleading
results when evaluating the sensitivity of an experiment to new physics.
In our case new physics means $|r|>0$ or $\alpha_\gamma < 1$.
Since only $\alpha_\gamma$ is the experimental observable, an experiment will have
established new physics at the 3$\sigma$ level if it finds $(1-\alpha_\gamma)/
\sigma_{\alpha_\gamma} > 3$.
Evaluating the corresponding sensitivity for $|r|>0$ with Eq.~(\ref{eq:err_r})
(and Eq.~(\ref{eq:alphaLO})) would result in 
\begin{equation*} 
    \frac{|r|}{\sigma_{|r|}} = (1+\alpha_\gamma)  
    \frac{1-\alpha_\gamma}{\sigma_{\alpha_\gamma}} ,
\end{equation*}
a value that is too large by a factor of typically almost two!%
\footnote{We thank Yuehong Xie for bringing this point to our attention.}
We therefore only use $(1-\alpha_\gamma)/\sigma_{\alpha_\gamma}$ to estimate 
sensitivities.%
\footnote{The reader should be advised that the sensitivity estimates presented in 
Ref.~\cite{Hiller-Kagan} are based on Eq.~(\ref{eq:err_r}) and therefore suffer from the same problem.}

In the case of spin-1/2 $\Lambda$ baryons (Sec.~\ref{subsec:J=1/2}), $\alpha_\gamma$
will be extracted from a fit to a distribution of the type
$1-s_\gamma\cos\theta$ (Eqs.~\ref{eq:ang_gamma} and \ref{eq:ang_p}),
i.e., $\alpha_{\gamma}= s_\gamma/a$, where $a$ is either the $\Lambda_b$
polarization $P_{\Lambda_b}$ or the weak decay parameter $\alpha_{p,\frac{1}{2}}$.
The statistical error on $\alpha_\gamma$ is therefore
\begin{equation*}
\sigma_{\alpha_{\gamma}}=
\frac{1}{a}\sqrt{\alpha_{\gamma}^2\sigma_{a}^2+ \sigma^2_{s_\gamma}}
.
\end{equation*}
For the statistical error on the slope $s_\gamma$ expected for a fit to $N$ events
we use the empirical formula
\begin{equation*}
\sigma_{s_\gamma} = 1.752 \sqrt{\frac{1 - 0.71 \cdot s_\gamma^2}{N}}
\end{equation*}
obtained with a fast simulation tool \cite{RooFit}.

Similarly, the extraction of $\alpha_\gamma$ from spin-3/2 $\Lambda$ baryon decays
(Sec.~\ref{subsec:J=3/2}) proceeds via (cf.\ Eq.~(\ref{eq:alpha_gamma_1/2_3/2}))
\begin{equation*}
\alpha_{\gamma} = \frac{s_{\gamma}}{2P_{\Lambda_b}}
\left(1-\frac{3}{\alpha_{p, \frac{3}{2}}}\right)
\end{equation*}
with statistical error
\begin{equation*}
\sigma_{\alpha_{\gamma}} = \sqrt{\alpha_{\gamma}^2\left(
\frac{\sigma_{s_\gamma}^2}{s_\gamma^2}+\frac{\sigma_{P_{\Lambda_b}}^2}{P_{\Lambda_b}^2}\right)+
\frac{9}{4}~\frac{s_\gamma^2}{P_{\Lambda_b}^2}~\frac{\sigma^2_{\alpha_{p,\frac{3}{2}}}}
{\alpha_{p,\frac{3}{2}}^4}},
\end{equation*}
and the statistical error on the proton asymmetry parameter from $N$ events is approximated
by another empirical formula obtained from simulation,
\begin{equation*}
\sigma_{\alpha_{p,\frac{3}{2}}} =
\frac{3.48 - 3.16 \cdot \alpha_{p,\frac{3}{2}}}{\sqrt{N}}.
\end{equation*}

\section*{Acknowledgments}

We are indebted to Gudrun Hiller for very fruitful discussions.
We also thank Maurice Gailloud and Stefano Villa for
carefully reading the manuscript.
This work was supported by the Swiss National Science Foundation under
grant Nr.~620-066162.

\end{document}